\pdfoutput=1
\PassOptionsToPackage{bookmarks=false,colorlinks=true,linkcolor=blue,urlcolor=cyan,filecolor=magenta,citecolor=red,pdfstartview=FitV,hyperfootnotes=false,pdftitle={Carroll swiftons},pdfauthor={Florian Ecker, Daniel Grumiller, Marc Henneaux, Patricio Salgado-Rebolledo},pdfsubject={Carroll swiftons},pdfkeywords={Carroll swiftons},bookmarksopen=true}{hyperref}
\documentclass[aps,prl,10pt,twocolumn,superscriptaddress]{revtex4}

\usepackage{mathrsfs, amssymb, amsmath, amsfonts, txfonts, latexsym, graphicx, orcidlink, accents} 

\newcommand{\eq}[2]{\begin{equation} #1 \label{#2} \end{equation}}
\DeclareMathOperator{\extdm}{d}
\newcommand{\extd}{\extdm \!}

\newcommand{\XP}{X_{\textrm{\tiny P}}}
\newcommand{\XH}{X_{\textrm{\tiny H}}}

\DeclareFontFamily{OT1}{rsfs}{}
\DeclareFontShape{OT1}{rsfs}{m}{n}{ <-7> rsfs5 <7-10> rsfs7 <10->rsfs10}{} 
\DeclareMathAlphabet{\mycal}{OT1}{rsfs}{m}{n}

\begin{document}


\title{Carroll swiftons}

\author{\orcidlink{0000-0002-0449-0081}Florian Ecker}
\email{fecker@hep.itp.tuwien.ac.at}
\affiliation{Institute for Theoretical Physics, TU Wien, Wiedner Hauptstrasse 8–10/136, A-1040 Vienna,
Austria}

\author{\orcidlink{0000-0001-7980-5394}Daniel Grumiller}
\email{grumil@hep.itp.tuwien.ac.at}
\affiliation{Institute for Theoretical Physics, TU Wien, Wiedner Hauptstrasse 8–10/136, A-1040 Vienna,
Austria}

\author{\orcidlink{0000-0002-8912-6384}Marc Henneaux}
\email{marc.henneaux@ulb.be}
\affiliation{Universit\'e Libre de Bruxelles and International Solvay Institutes, ULB-Campus Plaine CP231, B-1050 Brussels, Belgium}
\affiliation{Coll\`ege de France,  Universit\'e PSL, 11 place Marcelin Berthelot,  75005 Paris, France}

\author{\orcidlink{0000-0001-6832-6785}Patricio Salgado-Rebolledo}
\email{psalgadoreb@hep.itp.tuwien.ac.at}
\affiliation{Institute for Theoretical Physics, TU Wien, Wiedner Hauptstrasse 8–10/136, A-1040 Vienna,
Austria}

\date{\today}

\begin{abstract}
We construct Carroll-invariant theories with fields propagating outside the Carroll lightcone, i.e., at a speed strictly greater than zero (``Carroll swiftons''). We first consider models in flat Carroll spacetime in general dimensions, where we present scalar and vector Carroll swifton field theories. We then turn to the coupling to gravity and achieve in particular in two dimensions a Carroll invariant scalar swifton by coupling it suitably to Carroll dilaton gravity. Its backreaction on the geometry generates dynamical torsion. 
\end{abstract}
%
%

\maketitle


\section{Introduction}

Once relegated as the overlooked younger sibling of Galilean physics, Carrollian physics has surged into prominence, becoming ubiquitous over the past decade. Carroll symmetries can be obtained formally as the vanishing speed of light limit from Poincar\'e symmetries \cite{Levy1965,SenGupta1966OnAA}. As a consequence of this limit, space is absolute but time relative, and the lightcone collapses. It is partly due to these counterintuitive features  that Carroll symmetries were not appreciated by most of the physics community for quite a while. 

A decade ago, it was realized that Minkowski spacetime secretly is endowed with a Carroll structure: the metric at null infinity, $0\cdot\extd u^2+\extd\Omega^2$, has Carroll signature, where $u$ is retarded (or advanced) time and $\extd\Omega^2$ the metric of the celestial sphere. Moreover, the asymptotic symmetries of asymptotically flat spacetimes  discovered by Bondi, van~der~Burgh, Metzner, and Sachs (BMS) \cite{Bondi:1962,Sachs:1962} match precisely with conformal Carroll symmetries \cite{Duval:2014uoa, Duval:2014uva, Duval:2014lpa}. See Refs.~\cite{Ciambelli:2018wre, Figueroa-OFarrill:2021sxz, Herfray:2021qmp,Mittal:2022ywl, Campoleoni:2023fug} for more on the Carroll structure at null infinity. It was soon realized that also generic null hypersurfaces have a Carroll structure \cite{Penna:2015gza,Penna:2018gfx,Donnay:2019jiz,Ciambelli:2019lap,Redondo-Yuste:2022czg,Freidel:2022vjq,Gray:2022svz,Ciambelli:2023mir,Ciambelli:2023mvj}. 

The ubiquity of null hypersurfaces in physics explains the rapid increase of research that exploits Carrollian symmetries. Among the most prominent ones is the Carroll approach to flat space holography in three \cite{Barnich:2006av,Bagchi:2010zz,Bagchi:2012yk,Barnich:2012xq,Bagchi:2012xr,Barnich:2012rz,Bagchi:2013lma,Bagchi:2014iea,Barnich:2015mui,Campoleoni:2015qrh,Bagchi:2015wna,Bagchi:2016bcd,Jiang:2017ecm,Grumiller:2019xna,Apolo:2020bld} and four dimensions \cite{Ciambelli:2018wre,Figueroa-OFarrill:2021sxz,Herfray:2021qmp,Donnay:2022aba,Bagchi:2022emh,Campoleoni:2022wmf,Donnay:2022wvx,Mittal:2022ywl,Campoleoni:2023fug,Bagchi:2023fbj,Saha:2023hsl,Salzer:2023jqv,Nguyen:2023vfz,Nguyen:2023miw,Bagchi:2023cen,Mason:2023mti,Have:2024dff}. 

Other physics applications of Carroll symmetries include quantum gravity \cite{Isham:1975ur,Teitelboim:1978uc,Teitelboim:1978wv,Henneaux:1979vn,Henneaux:1981su}, tachyon condensates \cite{Gibbons:2002tv}, the fluid/gravity correspondence \cite{deBoer:2017ing,Ciambelli:2018xat,Campoleoni:2018ltl,Ciambelli:2020eba,Ciambelli:2020ftk,Freidel:2022bai,Petkou:2022bmz}, tensionless strings \cite{Bagchi:2015nca,Bagchi:2019cay,Bagchi:2020ats,Bagchi:2022iqb,Fursaev:2023lxq,Fursaev:2023oep}, cosmology \cite{Henneaux:1982qpq,Damour:2002et,deBoer:2021jej}, current-current deformations \cite{Rodriguez:2021tcz,Parekh:2023xms}, Hall effects \cite{Marsot:2022imf}, fractons \cite{Bidussi:2021nmp,Figueroa-OFarrill:2023vbj,Figueroa-OFarrill:2023qty,Pena-Benitez:2023aat,Perez:2023uwt}, flat bands \cite{Bagchi:2022eui}, Bjorken flow \cite{Bagchi:2023ysc}, supersymmetry and supergravity \cite{Ravera:2019ize,Ali:2019jjp, Ravera:2022buz,Kasikci:2023zdn}, and black holes \cite{Penna:2018gfx,Donnay:2019jiz,Gray:2022svz,Redondo-Yuste:2022czg,Bicak:2023rsz,Ecker:2023uwm,Bagchi:2023cfp}. Moreover, it is natural to gauge the Carroll algebra \cite{Hartong:2015xda} and establish Carroll gravity theories \cite{Henneaux:1979vn,Bergshoeff:2017btm,Ciambelli:2018ojf,Matulich:2019cdo,Grumiller:2020elf,Gomis:2020wxp,Perez:2021abf,Hansen:2021fxi,deBoer:2021jej,Concha:2021jnn,Figueroa-OFarrill:2022mcy,Campoleoni:2022ebj,Miskovic:2023zfz}, which may exhibit Carroll black hole solutions \cite{Ecker:2023uwm}. See \cite{deBoer:2023fnj,Ciambelli:2023xqk} for more Refs.

To better understand Carrollian field theories (the purported duals to asymptotically flat spacetimes), it is necessary to construct simple (but non-trivial) examples of such field theories. The Carrollian algebra in the context of scalar fields and particles was studied in \cite{Bergshoeff:2014jla,Henneaux:2021yzg,Bagchi:2022eav,Bekaert:2022oeh,Bergshoeff:2022eog,Rivera-Betancour:2022lkc,Ekiz:2022wbi,Baiguera:2022lsw,deBoer:2023fnj,Kasikci:2023tvs,Casalbuoni:2023bbh,Cerdeira:2023ztm,Kamenshchik:2023kxi,Zhang:2023jbi,Ciambelli:2023tzb}. 

Two versions of a Carrollian scalar field theory have particularly drawn attention in the past. In one version (dubbed ``electric''), the variation of the action
\eq{
I_e = \tfrac{1}{2}\int\extd^n x \,(\partial_t\phi)^2
}{eq:cips7}
establishes ultra-local equation of motion $\partial_t^2\phi=0$: the scalar field may depend on time but becomes ultra-local (i.e., devoid of spatial derivatives). In the other version (dubbed ``magnetic''), the variation of the action
\eq{
I_m = \int\extd^n x\, \Big(\pi\,\partial_t\phi - \tfrac{1}{2}\,\delta^{ij}\partial_i\phi\partial_j\phi\Big)
}{eq:cips9}
establishes the time-independence constraint $\partial_t\phi=0$ together with a Laplace equation of motion, $\delta^{ij}\,\partial_i\partial_j\phi=0$. Neither of these cases leads to scalar fields propagating with a finite, non-vanishing velocity. 

One might expect such behavior since the Carrollian collapse of the lightcone and the absoluteness of space suggest fractonic behavior, i.e., ``nothing can move''.  Carroll causality forces information to stay within the lightcone, i.e., to remain at the same spatial location. Propagation at a non-vanishing speed $v>0=c_{\textrm{\tiny Carroll}}$ would define a tachyonic-like behavior. 
 
It has been argued recently (see, e.g., \cite{deBoer:2023fnj}) that Carroll tachyons have their physical place and interest. Furthermore, given the existence of Carroll black hole thermodynamics \cite{Ecker:2023uwm}, one could hope that there are (scalar) modes as carriers of (some Carrollian analog of) Hawking quanta. Actions for Carrollian tachyonic particles have been considered in Refs.~\cite{deBoer:2021jej,Casalbuoni:2023bbh,deBoer:2023fnj}. 

The main goal of this Letter is to construct and discuss examples of Carroll-invariant actions for (interacting) fields allowing propagation at a non-vanishing velocity in arbitrary dimensions, both with and without (Carroll) gravity. We restrict our attention to scalar and electromagnetic fields.

Contrary to Lorentz tachyons, which usually come with pathologies associated with the unboundedness of their energy, we explicitly show that the energy is bounded from below, at least for two of our models (we have not investigated the other similar models).  This suggests that these models are free from the standard Lorentz tachyonic instabilities. Therefore, we refrain from referring to the excitations of our fields as ``tachyons''. Instead, we use the translated expression ``swiftons'' to highlight simultaneously the propagation outside the lightcone and the absence of tachyonic instabilities.

\section{Carroll spacetimes}\label{sec:2}

We first review, in this section and the next, selected elements of Carroll spacetimes, their symmetries, and Carroll gravity.  This is done partly to make this Letter more accessible to people outside the field and partly to fix the notation. However, this is not a complete review nor a full introduction, for which we refer instead to Refs.~\cite{Henneaux:1979vn,Duval:2014uoa,Bergshoeff:2014jla,Hartong:2015xda,Bergshoeff:2017btm,Bagchi:2019xfx,Ecker:2023uwm}.

Carroll spacetimes in $n=d+1$ dimensions have a degenerate signature $(0,+,\ldots,+)$ with $d$ pluses.  To characterize such spacetimes, in addition to a Carroll metric $h_{\mu\nu}$, we need a Carroll vector $v^\mu$ in the kernel of the metric, $v^\mu\,h_{\mu\nu}=0$. This is equivalent to endowing the manifold with a non-vanishing volume element $\Omega$ \cite{Henneaux:1979vn}.  The Carroll analog of the Minkowski metric is given by the Carroll metric $h_{\mu\nu}=\delta_{ij}\,\delta^i_\mu\,\delta^i_\nu$ and the vector field $v=v^\mu\partial_\mu=\partial_t$, implying the unit volume $\Omega=1$. Below, we sometimes use the abbreviation $\dot\phi\equiv{v}^\mu\partial_\mu\phi$.

While the metric $h_{\mu\nu}=\delta_{ij}\,\delta^i_\mu\,\delta^i_\nu$ is invariant under all Carroll transformations,
its ``inverse metric'' $h^{\mu\nu}\equiv\delta^{ij}\,\delta_i^\mu\,\delta_i^\nu$ is not invariant under Carroll boosts.  This means that one cannot raise tensor indices in an invariant way.  However, for ``transverse'', or ``spacelike'' co-vectors $\theta_\mu$, defined as being orthogonal to $v^\mu$, $v^\mu\theta_\mu=0$, the norm squared $\theta_\mu\theta^\mu$ is well-defined and positive.  The same is true for more general covariant tensors $\theta_{\mu_1\mu_2\cdots\mu_k}$ transverse on all their indices.

In terms of the Carroll structure, the electric action \eqref{eq:cips7} can be written in a manifestly invariant form as $I_e=\tfrac{1}{2}\int\extd^nx\,(v^\mu\partial_\mu\phi)^2$. The manifestly Carroll invariant form of the magnetic action \eqref{eq:cips9} is more involved and was worked out in \cite{Henneaux:2021yzg}.  The difficulty is that 
the term $\partial_\mu\phi\partial_\nu\phi$ is well-defined and invariant under all Carroll transformations, including boosts, only if $\partial_\mu\phi$ is transverse, i.e., $v^\mu\partial_\mu\phi=0$.  This is why the magnetic Carroll scalar action \eqref{eq:cips9} implements the constraint $v^\mu\partial_\mu\phi=0$, but this is only an on-shell relation. To write the action, one must go off-shell where $v^\mu\partial_\mu\phi$ is not zero, which necessitates introducing an extra auxiliary field  \cite{Henneaux:2021yzg,deBoer:2021jej}. 

This observation also explains why it is challenging to construct Carroll-invariant actions involving simultaneously time derivatives and spatial gradients of a single scalar field.

\section{Vielbein and connection}\label{sec:3}

It is often more convenient to work with Cartan-like variables, the temporal einbein $\tau=\tau_\mu\,\extd x^\mu$ (alias ``Ehresmann connection''), and the spatial vielbein, $e^a=e^a_\mu\,\extd x^\mu$, where $a$ is a spatial tangent space index, lowered by $\delta_{ab}$, with $a,b=1{\ldots}d$. The spacetime indices $\mu,\nu$ have the range $0{\ldots}d$ and the spatial indices $i,j$ the range $1{\ldots}d$. The Carroll metric is the bilinear of the spatial vielbein, $h_{\mu\nu}=e^a_{\mu}e^b_\nu\delta_{ab}$. The Carroll vector is the dual of the temporal einbein, $v^\mu\tau_\mu=-1$, and orthogonal to the spatial vielbein, $v^{\mu}e_\mu^a=0$. It is convenient to also introduce the inverse spatial vielbein, $e^\mu_a$, dual to the vielbein, $e^a_{\mu}e^\mu_b=\delta^a_b$, and orthogonal to the temporal einbein, $e^\mu_a\tau_\mu=0$.

In addition to the vielbein variables, we need connections. The connection associated with spatial rotations behaves in the same way as for Lorentzian or Galilean theories, so we do not review its properties. The Carroll boost connection, $\omega_a=\omega_{a\,\mu}\,{\extd}x^\mu$, is the gauge field associated with local Carroll boosts, which are abelian: $\delta_{\lambda}\omega_a=\extd\lambda_a$. The vielbein fields transform under Carroll boosts,
\eq{
\delta_\lambda\tau_\mu=-\lambda_a\,e^a_\mu\qquad\delta_\lambda e_a^\mu=-\lambda_a\,v^\mu\qquad\delta_\lambda e^a_\mu=0=\delta_\lambda v^\mu
}{eq:cips10}
in agreement with the transformation properties of the metric variables, $\delta_\lambda{v}^\mu=0=\delta_\lambda{h}_{\mu\nu}$.

\section{Bi-scalar model}\label{sec:6}

Our first key result of this Letter is the construction of Carroll invariant actions for scalar and electromagnetic swiftons, i.e., allowing propagation outside the Carroll lightcone.  

We start with a model that couples two scalar fields $\phi,\chi$ with canonically normalized kinetic terms and coupling constant $g$, thereby generalizing the model introduced in \cite{Baig:2023yaz} (see their Eq.~(2.1) as well as \cite{Kasikci:2023tvs}) to any Carroll background. Its action in covariant form 
\eq{
\boxed{\phantom{\Big(}
I_M = \tfrac12 \int\extd^nx \,\Omega\,\left(\left(v^\mu \partial_\mu \phi \right)^2 + \left(v^\mu \partial_\mu \chi \right)^2 + g\, B_\mu B^\mu\right)
\phantom{\Big)}}
}{eq:cips01}
contains the manifestly transverse covariant vector
\eq{
\boxed{\phantom{\Big(}
B_\nu \equiv v^\mu \left(\partial_\mu \phi \partial_\nu \chi - \partial_\mu \chi \partial_\nu \phi \right) \equiv 2 v^\mu \partial_{[\mu} \phi \partial_{\nu]} \chi\,.
\phantom{\Big)}}
}{eq:Bdef}
Because $B_\nu$ involves simultaneously time and spatial derivatives, the model allows propagation off the Carroll lightcone (see below). Antisymmetry of the coefficient of $v^\mu$ in $B_\nu$ is crucial for transversality, which would not hold if $\phi=\chi$ since then $B_\nu$ would be identically zero. By contrast, two distinct scalar fields can ``mutualize'' their derivatives in a non-trivial way through the identically transverse vector $B_\nu$. Each scalar field is crucial for its mutualistic partner in our construction.  

The Hamiltonian for this model can be derived straightforwardly, noting that the above Lagrangian density can be rewritten as [$\phi^A\equiv(\phi,\chi)$]
\begin{equation}
\mathcal L = \frac{\sqrt{h}}{2 N} H_{AB} \accentset{\circ}{\phi}^A \accentset{\circ}{\phi}^B \quad \,\;\; H_{AB} = \begin{pmatrix} 1+  g (\partial \chi)^2& - g\partial\phi\cdot\partial\chi \\ - g\partial\phi\cdot\partial\chi & 1+  g (\partial \phi)^2\end{pmatrix}\,.
\end{equation}
Here, $N$ is the Carroll lapse, $h$ is the determinant of the spatial metric $h_{mn}$ with inverse $h^{mn}$, $\partial\phi^A\cdot\partial\phi^B \equiv h^{mn} \partial_m \phi^A \partial_n \phi^B$ and $\accentset{\circ}{\phi}^A \equiv \dot{\phi}^A - N^k \partial_k \phi^A$ with $N^k$ the Carroll shift (see  \cite{Henneaux:1979vn}).

The inverse matrix $H^{AB}$ is $H^{AB}=\frac{1}{D}\left(\delta^{AB}+g(\partial\phi^A)\cdot(\partial\phi^B)\right)$ where $D$ is the determinant $D=1+g(\partial\phi)^2+g(\partial\chi)^2+g^2\left( (\partial\phi)^2(\partial \chi)^2-(\partial\phi\cdot\partial\chi)^2\right)$, which obeys $D\geq1$ for $g\geq0$, implying that the field space metric $H_{AB}$ has Euclidean signature.

The Hamiltonian is $N \mathcal H + N^k \mathcal H_k$ where the momentum density is $\mathcal H_k = \pi_A \partial_k \phi^A$ and  the energy density 
\begin{equation}
\mathcal H = \frac{1}{2\sqrt{h}}\, H^{AB}\, \pi_A \pi_B
\end{equation}
is bilinear in the conjugate momenta $\pi_A$. Since the quadratic form $H^{AB}\,\pi_A\pi_B$ is positive definite, the energ density is bounded from below by zero.

An instructive and not completely trivial computation shows that the Poisson brackets of the energy densities at different spacelike points vanish, $\{\mathcal{H}(x),\mathcal{H}(x')\}=0$, in agreement with the general argument of \cite{Henneaux:2021yzg}.  This ensures that the constraints $\mathcal H^T \approx 0$, $\mathcal H^T_k \approx 0$ of the dynamical gravity $+$ matter system
\begin{equation}
I = \int d^n x \left(\pi^{ij} \dot{h}_{ij} + \pi_A \dot{\phi}^A - N\mathcal H^T - N^k\mathcal H_k^T \right)
\end{equation}
are first class as they should.  In the above action, $\pi^{ij}$ are the conjugate momenta to the spatial metric while $\mathcal H^T = \mathcal H^G + \mathcal H $ and $\mathcal H^T_k = \mathcal H^G_k + \mathcal H_k$ are the sums of the Carroll gravity (in either the electric or magnetic version) and matter contributions to the Hamiltonian and momentum constraints.

The equations of motion following from the mutualistic scalar action \eqref{eq:cips01}
\begin{subequations}
\label{eq:cips3}
\begin{align}
\partial_\mu\Big(\Omega\,\big(v^\mu v^\nu\partial_\nu\chi-gh^{\alpha\nu}B_\nu v^\mu\partial_\alpha\phi+gh^{\mu\nu}B_\nu v^\alpha\partial_\alpha\phi\big)\Big)&=0\\
\partial_\mu\Big(\Omega\,\big(v^\mu v^\nu\partial_\nu\phi+gh^{\alpha\nu}B_\nu v^\mu\partial_\alpha\chi-gh^{\mu\nu}B_\nu v^\alpha\partial_\alpha\chi\big)\Big)&=0
\end{align}
\end{subequations}
are coupled non-linear partial differential equations, and we have not tried to devise a general method to solve them.

\section{Taylor expanded Swiftons}

To gain insight into the theory, we use perturbative methods. A soluble case arises when one of the scalar fields, say $\chi=\chi_{\textrm{\tiny BG}}+{\cal{O}}(\epsilon^2)$, is a background field in addition to the geometric background (which we assume to be static) and the other one, $\phi=\epsilon\,\varphi$, is a small fluctuation on top ($\epsilon\ll1$). To leading order we obtain from \eqref{eq:cips3} the background solution
\eq{
\chi_{\textrm{\tiny BG}} = \chi_0(x^i) + \chi_1(x^i)\,t
}{eq:cips4}
where we used adapted coordinates such that $v=f(x^i)\,\partial_t$. For simplicity, we set $\chi_0=0$ and $\chi_1=1$.

Inserting this solution back into the mutualistic scalar action \eqref{eq:cips01} yields a quadratic action
\eq{
I[\varphi]=\tfrac12 \int\extd^nx\,\Omega\,\Big(\epsilon^2\,\big(v^\mu\partial_\mu\varphi\big)^2 + \epsilon^2\,g\,h^{\mu\nu}\,(\partial_\mu\varphi)(\partial_\nu\varphi)  + 1\Big)
}{eq:cips5}
for the fluctuations $\varphi$ on such a background. 

To get hyperbolic equations of motion, we need a negative coupling constant $g$, which may seem at odds with positivity of energy density. However, even for negative $g$, the energy density remains positive as long as the coupling constant obeys the inequality $g>-1/[(\partial\phi)^2+(\partial\chi)^2]$. In our perturbative context where both $(\partial\phi)^2$ and $(\partial\chi)^2$ are small the bound on the coupling constant is very weak.

Up to a cosmological constant term and conventions, the action \eqref{eq:cips5} coincides precisely with the one proposed in \cite{Ciambelli:2023xqk}, i.e., it combines electric \eqref{eq:cips7} and magnetic \eqref{eq:cips9} Carroll scalar field actions to a single electromagnetic one. However, as opposed to \cite{Ciambelli:2023xqk}, the theory maintains Carroll boost invariance. 

\section{Multi-scalar models}

The bi-scalar model \eqref{eq:cips01} can easily be generalized by the same ``mutualization trick'' to a multi-scalar theory. For instance, with three scalars we have the action [$\phi^A\equiv(\phi,\chi,\psi)$]
\eq{
I_{M_3} = \tfrac12\int\extd^nx\,\Omega\, \Bigg( \sum_{A= 1}^3\left(v^\mu \partial_\mu \phi^A \right)^2  + g\, B_{\mu \nu}B^{\mu \nu}\Bigg)
}{eq:nolabel}
with the transverse tensor
\eq{B_{\mu \nu} \equiv v^\rho B_{\mu \nu \rho}\qquad \qquad  B_{\mu \nu \rho} \equiv \partial_{[\mu} \phi \partial_\nu \chi \partial_{\rho]} \psi\,.}
{eq:whatever}
The interaction term in \eqref{eq:nolabel} is of order six in the derivatives, but only quadratic in the time derivatives.

\section{Electromagnetic model}

A non-trivially interacting electromagnetic model can also be constructed using the same idea. Consider 
\begin{equation}
C_{\mu \nu \rho} \equiv v^\sigma C_{\mu \nu \rho \sigma} \qquad\qquad  C_{\mu \nu \rho \sigma} \equiv F_{[\mu \nu} F_{\rho \sigma]}
\end{equation}
where $F_{\mu\nu}=\partial_\mu{A}_\nu-\partial_\nu{A}_\mu$ is the electromagnetic field.  This is a transverse three-form, $C_{\mu\nu\rho}\nu^\rho=0$. Hence, again its square $C_{\mu\nu\rho}C^{\mu\nu\rho}$ is well-defined. 

The electromagnetic Carroll swifton action
\eq{
\boxed{\phantom{\Big(}
I_{\textrm{\tiny EM}} =\tfrac12 \int\extd^nx\,\Omega\,\Big(\big(v^\mu F_{\mu \nu} \big)^2 + g\, C_{\mu \nu \rho}C^{\mu \nu \rho}\Big)
\phantom{\Big)}}
}{eq:EM}
yields the equations of motion
\eq{
\partial_\lambda\Big(\Omega\big(v^\mu v^{[\sigma}h^{\lambda]\nu}+gv^\nu C^{\sigma\lambda\mu}+gv^{[\lambda}C^{\sigma]\mu\nu}\big)F_{\mu\nu}\Big)=0
}{eq:EMeom}
that are again coupled non-linear partial differential equations. 

The energy density is given by a similar expression as in the scalar case
\begin{align}
    \mathcal{H}=\frac{1}{2\sqrt{h}}H_{ij}\pi ^i\pi ^j 
\end{align}
where $H_{ij}$ is the inverse of the symmetric matrix
\begin{align}
    H^{ij}=h^{ij}+\frac{2g}{3}\Big(h^{il}h^{mk}h^{nj}+\frac{1}{2}h^{ij}h^{lk}h^{mn}\Big)F_{lm}F_{kn} ~.
\end{align}
In four dimensions the determinant of the latter is given by $D=h^{-1}(1+2gB^2/3)$ with $B^2=h^{ik}h^{jl}F_{kl}F_{ij}/2$ which makes $\mathcal{H}$ positive definite if $g>-3/(2B^2)$, allowing in particular again negative values of $g$ in a perturbative context.

For the special case of a four-dimensional flat Carroll background we linearize the equations of motion \eqref{eq:EMeom} with a constant electric background field plus fluctuations on top,
\begin{equation}
F_{ij}
\equiv \epsilon\,\mathcal B_{ij}\qquad\qquad F_{ti}= -\delta_i^x\,E-\epsilon\,\mathcal E_i \,.
\end{equation} 
The linearized equations of motion are solved by plane waves 
\eq{
\mathcal E_i = A_i e^{i(k_yy+k_zz-\omega t)}\qquad\mathcal{B}_{ij}=B\,\big(\delta_i^y\delta_j^z-\delta_i^z\delta_j^y\big)e^{i(k_yy+k_zz-\omega t)}
}{eq:planewave}
subject to the dispersion relation $\omega^2=c_{\textrm{\tiny{eff}}}^2\,(k_y^2+k_z^2)$ with the effective speed of light $c_{\textrm{\tiny{eff}}}^2=-\frac{2g}{3}\,E^2$, the transversality condition $k_yA_y+k_zA_z=0=A_x$, and the normalization $B^2=A_y^2+A_z^2$. As in the bi-scalar model, we need negative $g$ to have hyperbolic equations of motion with a real effective propagation speed for the swiftons.

Denoting $J_1=(v^\mu{F}_{\mu\nu})^2$ and $J_2=C_{\mu\nu\rho}C^{\mu\nu\rho}$, this model can be generalized to Lagrange densities of the form $\mathcal{L}=\Omega\,f(J_1,J_2)$. It would be interesting to determine which choice of $f$ leads to the duality-invariant theory constructed in \cite{Bunster:2012hm}.

The electromagnetic model can be directly coupled to the bi-scalar action \eqref{eq:cips01}. Indeed, in terms of the complex field $\Phi=\phi+i\chi$ the mutualistic Lagrangian density takes the simple form $\mathcal L_M=\frac12 \dot\Phi\dot\Phi^* + \frac{g}{8}h^{ij}|\dot\Phi \partial_i\Phi^*-\dot\Phi^*\partial_i\Phi |^2$. Thus, the global $U(1$) symmetry of the model can be made local by introducing standard minimal coupling $\partial_\mu\rightarrow \partial_\mu-iA_\mu$. 

\section{Scalars propagating on Carroll black holes}\label{sec:4}

We now focus on two spacetime dimensions and construct swiftons coupled to gravity. The reason for considering two dimensions (2d) is that all known Carroll black hole solutions are described by 2d models (intrinsically or by dimensional reduction). Below all tangent space indices $a,b$ are dropped since we have only one spatial dimension.

Carroll dilaton gravity in 2d was introduced in \cite{Grumiller:2020elf} (see also \cite{Gomis:2020wxp}). Its action
\eq{
I_{\textrm{\tiny CDG}}\sim\int\big(X\extd\omega+\XH\,(\extd\tau+\omega\wedge e)+\XP\,\extd e+\tau\wedge e\,{\cal V}(X,\,\XH)\big)
}{eq:cips11}
depends on the temporal einbein $\tau$, the spatial einbein $e$, the Carroll boost connection $\omega$, the dilaton $X$, the Lagrange multiplier $\XH$ for the torsion constraint, and the Lagrange multiplier $\XP$ for the intrinsic torsion constraint. The potential function ${\cal V}(X,\XH)$ depends on Carroll boost-invariant scalar fields. The remaining scalar field $\XP$ is paramount for our construction below, as it transforms non-trivially under local Carroll boosts, 
\eq{
\delta_\lambda\XP=\lambda\,\XH\,.
}{eq:cips12}

All classical solutions for all models \eqref{eq:cips11} were constructed in \cite{Ecker:2023uwm}. Among them are Carroll black holes, which have thermal properties similar to Lorentzian black holes. To verify whether or not there is a Hawking-like effect \cite{Ankitprep}, we need to couple matter to Carroll black holes since otherwise, the theory has no local propagating degrees of freedom. This provides an additional motivation for considering Carroll swifton scalars propagating on Carroll black hole backgrounds.

The second key result of this Letter is the 2d Carroll swifton scalar field action,  
\eq{
\boxed{\phantom{\Big(}
I_{\textrm{\tiny 2d}}=\tfrac12\int\extd^2 x\,\Omega\, F\,\Big(\dot\phi^2 + g\,(\hat\partial\phi)^2 + h\,\dot\phi\,\hat\partial\phi\Big)
\phantom{\Big)}}
}{eq:cips15}
where the coupling functions $F,g,h$ may depend on the dilaton $X$ and the Carroll boost-invariant scalar $\XH$. The volume form is given by $\extd^2x\,\Omega=\tau\wedge{e}$. The most general Carroll invariant second order action \eqref{eq:cips15} combines non-trivially time- and space-derivatives of $\phi$ and has the merit of not introducing any extra structure besides the Carroll background. 
Here, we have defined the Carroll boost-invariant derivative 
\eq{
\boxed{\phantom{\Big(}
\hat\partial\equiv{e}^\mu\partial_\mu+\frac{\XP}{\XH}\,v^\mu\partial_\mu\,.
\phantom{\Big)}}
}{eq:derivative}
The last term in \eqref{eq:cips15} does not generalize to higher dimensions \cite{Bagchi:2022eav}, whereas the first two terms do. We stress that we have added a term in $\hat\partial$ that transforms like a St\"uckelberg field \cite{Hartong:2015xda}, but using only fields that were there already in the gravity action \eqref{eq:cips11}. 

The definition \eqref{eq:derivative} introduces the important restriction $\XH\neq0$, i.e., we are not allowed to sit on a Carroll extremal surface \cite{Ecker:2023uwm}. This requirement is the Carrollian pendant of considering the outside region of a black hole, so for most applications, $\XH\neq0$ is obeyed. If one needs to extend the action \eqref{eq:cips15} onto a Carroll extremal surface $\XH=0$, one can do so, for instance, by choosing $g\propto\XH^2$.

\section{Regge--Wheeler-type equation}

We provide as a pertinent example a scalar field propagating on a Carroll--Schwarzschild background (see, e.g., Eqs.~(213)-(219) in \cite{Ecker:2023uwm}), where we assume non-minimal coupling to the dilaton, $F=X\equiv{r}^2$, set $h=0$, and leave $g$ constant. Defining $\phi=\psi/r$ and the tortoise coordinate $r_\ast\equiv{r}+2m\ln(\frac{r}{2m}-1)$ yields the Regge--Wheeler-type equation
\eq{
\partial_t^2\psi+g\,\partial_{r_\ast}^2\psi=\frac{2gm}{r^3}\,\bigg(1-\frac{2m}{r}\bigg)\,\psi
}{eq:SBH}
where $m>0$ is the mass of the Carroll-Schwarzschild black hole, and we assume $r>2m$ to be outside the Carroll extremal surface. For the value of the coupling constant $g=-1$, the equation \eqref{eq:SBH} is identical to the s-wave sector of the Regge--Wheeler equation, see e.g.~\cite{Grumiller:2022qhx}, and thus standard results apply to this case. More generally, for negative (positive) $g$, the swifton equation \eqref{eq:SBH} is hyperbolic (elliptic).

\section{Dynamical torsion from scalar swiftons}\label{sec:5}

Without backreaction, our matter action \eqref{eq:cips15} (with $F=1$, $h=0$, and $g=\textrm{const.}$) is indistinguishable from the one in \cite{Ciambelli:2023xqk}. This ceases to be the case when taking into account backreactions of the scalar field on the Carroll geometry. 

To show this, we add it to the gravity action \eqref{eq:cips11} and consider here two of the field equations, namely the ones coming from variations with respect to $\XP$ and $\XH$:
\eq{
    \extd e =-\frac{g}{\XH}\,\dot\phi\,\hat\partial\phi\,\tau\wedge e\qquad
    \extd\tau + \omega\wedge e = \frac{g\XP}{\XH^2}\,\dot\phi\,\hat\partial\phi\,\tau\wedge e 
}{eq:angelinajolie}
The left-hand sides are, respectively, intrinsic torsion (so named because it cannot be altered by changing the connection) and standard torsion. The right-hand sides are zero without matter but are sourced by the same term $\dot\phi\,\hat\partial\phi$ in the presence of swifton matter. Therefore, if this expression is non-zero, then the backreactions of a scalar field that propagates on a Carroll background induce dynamical torsion (both intrinsic and standard). It would be of interest to explore further the dynamical properties of this theory.

\section{Discussion}\label{sec:7}

Apart from the duality-invariant electromagnetic Lagrangian of \cite{Bunster:2012hm}, there was to our knowledge no field theoretical Carroll-invariant Lagrangian that was neither of electric or magnetic type and allowed propagation off the Carrollian lightcone on arbitrary Carroll backgrounds.  

We have constructed and discussed in this Letter new scalar and vector models with this property. We have also explicitly verified for two of these models that Carroll tachyonic behavior is compatible with the positivity of the energy: Carroll tachyons need not be plagued by the pathologies of their Lorentz-invariant analogs, which is why we refer to them as ``Carroll swiftons''. 

We considered our models intrinsically Carrollian rather than as induced at some null hypersurface in a Lorentzian spacetime. In such a context, swiftons would correspond to Lorentzian tachyons in the ambient spacetime with the associated difficulties.

We conclude by mentioning further generalizations and applications. It seems worthwhile generalizing our scalar and vector swifton models to higher spin, half-integer spin, and supersymmetric swiftons. For applications in flat space holography, it would be gratifying to construct conformal swifton models since such models can be candidates for the field theory dual of asymptotically flat gravity theories. For the swifton models considered in our Letter, it could be rewarding to go beyond Taylor expansions and find non-perturbative swifton solutions while obeying the positivity constraint on energy density. Finally, it should be fruitful to study backreactions in (semi-)classical Carroll gravity with swiftons.



\bigskip

\begin{acknowledgments}
We thank Ankit Aggarwal, Arjun Bagchi, Rudranil Basu, Emil Have, Jelle Hartong, Niels Obers, and Dima Vassilevich for discussions and/or collaborations on related topics. We are grateful to Andreas Karch for pointing out their flat Carrollian version of the bi-scalar model \eqref{eq:cips01} in the context of spacetime subsystem symmetries.

The work of FE, DG, and PS-R was supported by the Austrian Science Fund (FWF), projects P 32581, P 33789, and P 36619. The work of MH was partially supported by  FNRS-Belgium (conventions FRFC PDRT.1025.14 and IISN 4.4503.15), as well as by funds from the Solvay Family. FE, DG, and PS-R acknowledge support by the OeAD travel grant IN 04/2022 and thank Rudranil Basu for hosting them at BITS Pilani in Goa in February 2024 through the grant DST/IC/Austria/P-9/202 (G).
\end{acknowledgments}


\providecommand{\href}[2]{#2}\begingroup\raggedright\endgroup

\end{document}